\documentclass[preprint,prb,superscriptaddress,amssymb,amsmath, floatfix,citeautoscript]{revtex4}

\usepackage[pdftex]{graphicx}

\newcommand{\p}{\partial }
\newcommand{\w}{\omega}
\renewcommand{\k}{\mathbf{k}}
\newcommand{\kh}{\hat{k}}

\newcommand{\gk}{\gamma_{\k}}

\newcommand{\Tzt}{T(z,t)}
\newcommand{\vzk}{v_{z\k}}

\newcommand{\tk}{\tau_{\k}}
\newcommand{\lk}{\Lambda_{z\k}}

\begin{document}

\title{Phonon heat conduction in layered anisotropic crystals}

\author{A. J. Minnich}
\email{aminnich@caltech.edu}
\affiliation{Division of Engineering and Applied Science\\
California Institute of Technology\\
Pasadena, CA 91125}

\begin{abstract}
The thermal properties of anisotropic crystals are of both fundamental and practical interest, but transport phenomena in anisotropic materials such as graphite remain poorly understood because solutions of the Boltzmann equation often assume isotropy. Here, we extend an analytical solution of the Boltzmann equation to highly anisotropic solids and examine its predictions for graphite. We show that the phonon mean free paths in the cross-plane direction can be comparable to those in the in-plane direction despite the low cross-plane thermal conductivity, which instead arises primarily from the differences in group velocities and phonon frequencies supported along each direction. Additionally, we demonstrate a method to reconstruct the anisotropic mean free path spectrum of crystals with arbitrary dispersion relations without any prior knowledge of their harmonic or anharmonic properties using observations of quasiballistic heat conduction.

\end{abstract}

\maketitle

\section{Introduction}

Thermal transport in anisotropic crystals is of both fundamental and practical importance. Anisotropic materials can possess extreme values of thermal conductivity that are difficult to achieve in isotropic solids, with the most familiar example being graphite with an exceptional thermal conductivity of 2000 W/mK in the basal plane but only 6 W/mK in the cross-plane \cite{touloukian_thermophysical_1970}. Graphene can be regarded as the extreme of anisotropy and has attracted tremendous attention \cite{jang_thickness-dependent_2010,ghosh_dimensional_2010, seol_two-dimensional_2010, bonini_acoustic_2012,lindsay_phonon_2014,paulatto_anharmonic_2013, bae_ballistic_2013,xu_length-dependent_2014}. The high in-plane thermal conductivity of graphene and graphite  makes them of considerable interest for heat spreading applications \cite{yan_graphene_2012}. At the other extreme, exceptionally low thermal conductivity has been reported in a variety of nanolaminates in the cross-plane direction \cite{chiritescu_ultralow_2007,losego_ultralow_2013}, and SnTe was recently reported to possess a record high thermoelectric figure of merit due largely to the low thermal conductivity along the b-axis \cite{zhao_ultralow_2014}. Anisotropic materials can also exhibit a number of physical effects that do not occur in isotropic materials such as phonon focusing due to non-spherical wavefronts \cite{von_gutfeld_heat_1964}.

As a readily available and prototypical anisotropic crystal, the thermal properties of graphite have been extensively studied, starting with lattice dynamics \cite{al-jishi_lattice-dynamical_1982,nicklow_lattice_1972,bowman_low-temperature_1958,krumhansl_lattice_1953} and thermal conductivity measurements by Slack \cite{slack_anisotropic_1962}, Taylor \cite{taylor_thermal_1966}, and Tanaka and Suzuki \cite{tanaka_thermal_1972}. Other measurements on graphite and graphite intercalation compounds were reported by Dresselhaus et al\cite{issi_electronic_1983,heremans_thermoelectric_1981,boxus_low_1981,dresselhaus_intercalation_1981}. Among the conclusions of these early works were that the phonon mean free path (MFP) along the c-axis of graphite is on the order of a few nanometers at room temperature. Other anisotropic layered materials that have been recently studied experimentally include multilayered films  \cite{chiritescu_low_2008}, muscovite crystals \cite{hsieh_pressure_2009}, organoclay nanolaminates, \cite{losego_ultralow_2013} and WSe$_{2}$ disordered crystals \cite{chiritescu_ultralow_2007}.

Theoretically, Klemens and Pedrazza obtained expressions for the phonon thermal conductivity of graphite assuming a 2D Debye model \cite{klemens_thermal_1994}. This treatment predicted an in-plane thermal conductivity that was in good agreement with the experimental value, but this 2D model cannot account for cross-plane heat conduction. While graphene has been studied extensively using ab initio calculations \cite{bonini_acoustic_2012,lindsay_phonon_2014,paulatto_anharmonic_2013}, most other theoretical treatments of graphite have considered only the thermal boundary conductance between graphite and other materials \cite{chen_anisotropic_2013, duda_extension_2009,prasher_thermal_2008}. Chen et al.\ recently examined the thermal boundary conductance of anisotropic solids using an anisotropic Debye model and found that an isotropic model leads to a substantial overprediction of the cross-plane thermal conductance, among other inaccuracies \cite{chen_anisotropic_2013}. Wei et al. examined the thermal conductivity of graphite using lattice dynamics and molecular dynamics (MD), concluding that a negative correlation exists between in-plane bonding strength and cross-plane thermal conductivity. \cite{wei_negative_2013} Heat conduction along the c-axis of graphite has been studied recently by MD simulations, which indicated that the phonons MFPs are considerably longer than predicted by simple kinetic theory  \cite{wei_phonon_2014}. This prediction appears to be supported by experiment \cite{yang_phonon_2014, harb_c-axis_2012}.

While these works provide important insights into thermal transport in graphite, substantially more insight into the spatial and temporal dynamics of thermal transport can be obtained using the Boltzmann transport equation (BTE) \cite{GangBook}. This formalism requires much less computation than MD simulations yet enables the dynamics of thermal transport to be calculated over a wide range of length scales, unlike simple formulas for thermal boundary conductance and thermal conductivity. However, this equation is nearly always solved assuming an isotropic crystal with a spherical Brillouin zone, and thus at present many dynamic aspects of heat conduction in anisotropic crystals remain poorly understood due to a lack of suitable theoretical approach.

Here, we investigate heat conduction in graphite using a recently reported analytic solution of the BTE \cite{Hua:2014}. This solution, which is equivalent to the Green's function of the BTE, allows an analytic description of the response of an infinite crystal to a volumetric heat input with arbitrary temporal and 1D spatial profile. We find that the simple assumption of velocity anisotropy largely explains the thermal conductivity anisotropy in graphite rather than differences in scattering along each direction. In fact, our result shows that c-axis MFPs can be comparable to those along the ab-axis even though the c-axis thermal conductivity is much smaller than that of the ab-axis. Further, we use our solution to develop a reconstruction method that allows the anisotropic MFP spectrum to be obtained from observations of quasiballistic transport without any prior knowledge of the harmonic or anharmonic properties of the crystal.

The paper is organized as follows. First, in Sec.\ \ref{sec:general} we derive the general solution to the BTE assuming an infinite crystal is subject to volumetric heat generation with an arbitrary temporal and 1D spatial profile. We then derive simplifications of this solution to enable analytic insights and specialize the equations to graphite. In Sec.\ \ref{sec:results}, we examine the effect of velocity anisotropy on the thermal conductivity and the transient response of the anisotropic crystal to heat impulses oriented along different crystallographic directions. Finally, we use these results to develop a method that can reconstruct the MFP spectrum of a crystal with an arbitrary dispersion relation.

\section{Theory} \label{sec:general}

\subsection{General solution of the BTE} 
We begin by briefly reviewing the derivation of the transport solution to the BTE as presented by Hua and Minnich for an isotropic crystal \cite{Hua:2014} and generalized to a solid with an arbitrary dispersion relation by Vermeersch et al\cite{vermeersch_superdiffusive_2014}. The one-dimensional BTE under the relaxation time approximation for transport along the $z$ direction is given by:
\begin{equation} \label{eq:BTE_energy}
\frac{\partial g_{\k}}{\partial t}+ \vzk \frac{\p g_{\k}}{\p z} = -\frac{g_{\k}-g_0(\Tzt)}{\tau_{\k}}+Q_{\k},
\end{equation}
where $g_{\k} = \hbar \omega (f_{\k}(z,t,\k)-f_0(T_0) )$ is the desired deviational distribution function, $Q_{\k}(z,t)$ is the spectral volumetric heat generation, $\vzk$ is the component of the phonon group velocity along the $z$ direction, $\tk$ is the phonon relaxation time, and $\k$ is the phonon wavevector in phase space.   $g_0(\Tzt)$ is the local equilibrium deviational distribution function and is given by:
\begin{equation}
g_0(\Tzt) = \hbar \w (f_{BE}(\Tzt) - f_0(T_0) ) \approx C_{\k} \Delta \Tzt.
\label{eq:BEDist_Linearized}
\end{equation}
assuming a small temperature rise $\Delta \Tzt = \Tzt - T_0$ relative to a reference temperature $T_{0}$. Here, $\hbar$ is the reduced Planck constant, $\w(\k)$ is the phonon frequency given by the dispersion relation, $f_{BE}(\Tzt)$ is the local Bose-Einstein distribution, $f_{0}(T_{0})$ is the Bose-Einstein distribution at $T_{0}$, and $C_{\k} = N_{pol }k_{B} (\chi/\sinh(\chi))^{2}$ is the mode specific heat, where $\chi=\hbar \w/2 k_{B} \Tzt$ and $N_{pol}$ is the number of polarizations. To close the problem, energy conservation is used to relate $g_{\k}$ to $\Delta \Tzt$ as 
\begin{equation}
\sum_{\k} \left[\frac{g_{\k}}{\tk}-\frac{C_{\k}}{\tk} \Delta \Tzt \right] = 0,
\label{eq:EnergyConservation}
\end{equation}
where the sum is performed over all phonon modes in the Brillouin zone.

An analytic solution to the above equations is obtained by performing a Fourier transform over $z$ and $t$ in Eq.\ \ref{eq:BTE_energy}, substituting the result into Eq.\ \ref{eq:EnergyConservation}, and solving for the temperature response $\Delta \widetilde{T}(\eta,\xi_{z})$, where $\eta$ and $\xi_{z}$ are the Fourier variables in the time and spatial domains, respectively. \cite{Hua:2014} This procedure assumes that the thermal transport occurs in an infinite domain so that a Fourier transform can be performed. The result is:

\begin{eqnarray}\label{eq:Temperature_FourierTransform}
\Delta \widetilde{T}(\eta,\xi_{z}) &=& \frac{ \sum_{\k} \widetilde{Q}_{\k} \gk^{-1} } { \sum_{\k} C_{\k} \tk^{-1} (1-\gk^{-1})}
\end{eqnarray}
where $\gk = 1 + i \eta \tk + i \xi_{z} \lk$ and $\lk = \vzk \tk$ is the component of the MFP along the $z$ direction. Once $\Delta \widetilde{T}$ is determined, $\widetilde{g}_{\k}$ can be expressed as:

\begin{equation} \label{}
\widetilde{g_{\k}}=(C_{\k}  \Delta \widetilde{T}+ \widetilde{Q}_{\k} \tk) \gk^{-1}
\end{equation}

The heat flux $\widetilde{q}_{z}$ along the $z$ direction is given by:
\begin{eqnarray} \label{}
\widetilde{q_{z}} &=& \frac{1}{V} \sum_{\k} \widetilde{g}_{\k} \vzk \\
\end{eqnarray}
where $V$ is the volume of the crystal. Solutions to each of these quantities in real-space can be easily obtained by inverse Fourier transform. This solution is valid for a crystal with an arbitrary dispersion relation. In the diffusive limit, these equations reduce to Fourier's law with a thermal conductivity given by: 

\begin{eqnarray} \label{eq:kz}
\kappa_{z} &=&  \frac{1}{V} \sum_{\k} C_{\k} \vzk^{2} \tk
\end{eqnarray}


\subsection{Weak quasiballistic regime} \label{sec:wqb}

We now derive a simplified solution to these equations under the assumption that $\eta \tau \ll 1$. This regime, previously denoted the weak quasiballistic regime by Hua and Minnich,\cite{Hua:2014} physically indicates that length scales are comparable to MFPs but the relevant timescales are much longer than relaxation times. This situation frequently occurs in thermal experiments such as transient grating spectroscopy for which grating periods $\lambda = 2 \pi / \xi_{z}$ are on a micron length scale but thermal decay times are tens to hundreds of nanoseconds, far longer than typical relaxation times \cite{johnson_direct_2013}.

The term $\gk^{-1}$ can be written as:
\begin{equation} \label{}
\gk^{-1} = \frac{1}{1 + i \eta \tk + i x} = \sum_{n=0}^{\infty} (-1)^{n} (i \eta \tk + i x)^{n}
\end{equation}
Expanding the polynomial and neglecting higher order terms in $\eta \tk$ yields:
\begin{equation} \label{}
\gk^{-1} = \sum_{n=0}^{\infty} (-ix)^{n} (1 - i(n+1) \eta \tk)
\end{equation}

These sums can be performed using the standard geometric series formula as:

\begin{equation} \label{}
1 - \gk^{-1} = \frac{i \eta \tk}{(1 + i x)^{2}} + 1 - \frac{1}{1 + i x}
\end{equation}

Substituting this result into the denominator of Eq. \ref{eq:Temperature_FourierTransform} gives:

\begin{equation} \label{eq:S1}
\sum_{\k} \frac{C_{\k}}{\tk} (1 - \gk^{-1}) = \sum_{\k} \frac{i \eta C_{\k}}{(1 + i x)^{2}} + \frac{q^{2}}{V} \sum_{\k} C_{\k} \vzk^{2} \tk \left[ \frac{1}{x^{2}} (1 - \frac{1}{1 + i x}) \right]
\end{equation}

Inserting this result into Eq. \ref{eq:Temperature_FourierTransform} and simplifying yields $\Delta T$ under the assumption $\eta \tau \ll 1$:

\begin{equation} \label{}
\Delta T = \frac{Q}{i \eta C + q^{2} V^{-1} \sum_{\k} \kappa_{z\k} (1 + x^{2})^{-1}}
\end{equation}

where the spectral thermal conductivity $\kappa_{z\k}= C_{\k} \vzk^{2} \tk$. To obtain this result we have neglected quasiballistic terms multiplying $i \eta C$ as in the weakly quasiballistic regime ballistic modes have a small contribution to the heat capacity. We also assume that $Q_{\k} \sim C_{\k}$, allowing us to use the same assumption to simplify the numerator. Finally, we have retained only the real part of Eq.\ \ref{eq:S1} as the imaginary part is odd and sums to zero.

Comparing this result with the Green's function of the heat equation, $\Delta T(\xi_{z}, \eta) =Q \times(i \eta C + q^{2} \kappa_{z})^{-1}$, we observe that in the weakly quasiballistic regime the effective thermal conductivity can be expressed as

\begin{equation} \label{eq:keff}
\kappa_{eff} = \frac{1}{V} \sum_{\k} \kappa_{z\k} S(x)
\end{equation}

where $S(x) = (1+x^{2})^{-1}$ is the suppression function. This result is identical to that derived previously by Hua and Minnich \cite{Hua:2014} with two important differences. First, this result is valid for a crystal with an arbitrary dispersion relation, unlike the prior result which is valid only for isotropic crystals. Second, the independent variable $x=\xi_{z} \lk$ contains the component of the MFP along the transport direction rather than the magnitude of the MFP as in the isotropic case.


\subsection{Application to graphite}

We now seek to use this solution to study thermal transport in highly anisotropic materials. In this work we focus on graphite-like layered crystals with a dispersion given by an anisotropic Debye model \cite{hsieh_pressure_2009, chen_anisotropic_2013}. While the Debye model has known shortcomings for transport calculations,\cite{AustinEES} for the present work it is sufficient to describe the key physics while enabling analytic expressions for transport quantities. In this model, the dispersion relation is given by:

\begin{equation} \label{}
\w^{2} = v_{ab}^{2} k_{ab}^{2} + v_{c}^{2} k_{c}^{2}
\end{equation}

where $v_{ab}$ and $k_{ab}$ are the sound velocity and wavevector along the $ab$ direction and $v_{c}$ and $k_{c}$ are the corresponding quantities for the $c$ direction. This form of the dispersion assumes that symmetry exists in two of the crystal directions such that $v_{a} = v_{b} = v_{ab}$. We take the $ab$ direction to represent the in-plane direction of graphite while $c$ represents the cross-plane direction. The phonon group velocity along crystal axis $j$ is given as 

\begin{equation} \label{}
v_{j} = \frac{\p \w}{\p k_{j}} = \frac{v_{j,s}^{2} k_{j}}{\omega(\k)}
\end{equation}

where $v_{j,s}$ is the sound velocity. 

The transport properties for a solid with this dispersion can be expressed as sums, as presented in Sec.\ \ref{sec:general}, over the phase space states contained in an ellipsoidal Brillouin zone specified by:
\begin{equation} \label{}
\frac{k_{ab}^{2}}{k_{ab,m}^{2}} + \frac{k_{c}^{2}}{k_{c,m}^{2}} = \kh_{ab}^{2} + \kh_{c}^{2} = 1
\end{equation}

where $k_{j,m}$ denote the maximum wavevectors in the Brillouin zone along crystal axis $j$ and $\kh_{j}$ are wavevectors normalized to $k_{j,m}$. These sums can be simplified by replacing them with integrals over the ellipsoidal Brillouin zone using the relation $V^{-1} \sum_{\k} = (2\pi)^{-3} \int d^{3} \k$. This integral can then be further simplified by transforming the volume of integration to a unit sphere. To do so, we define $k_{c}=\kh k_{c,m} \cos \theta$, $k_{a}=\kh k_{ab,m} \sin \theta \cos \phi$, and $k_{b}=\kh k_{ab,m} \sin \theta \sin \phi$, and $\kh^{2} = \kh_{ab}^{2} + \kh_{c}^{2}$. The resulting integral is then:

\begin{equation} \label{}
\frac{1}{V} \sum_{\k} b(\k) = A \int_{0}^{2\pi} \int_{-1}^{1} \int_{0}^{1} b(\kh, \mu, \phi) \kh^{2} d\kh d\mu d\phi 
\end{equation}

where $b$ is an arbitrary function of $\k$, $\mu=\cos \theta$, and $A= k_{c,m} k_{ab,m}^{2}/(2\pi)^{3}$.

Using this transformation, the temperature response $\Delta \widetilde{T_{j}}$ for heat input $\widetilde{Q}_{\k}(\eta, \xi_{z})$ when the $z$-axis is aligned with crystal axis $j$ is:

\begin{equation} \label{DTgraphite}
\Delta \widetilde{T_{j}}(\eta,\xi_{z}) = \frac{ \int_{0}^{1} \int_{-1}^{1} \widetilde{Q}_{\k} G_{j} \kh^{2} d\kh d\mu} { \int_{0}^{1} \int_{-1}^{1} C_{\k}\tk^{-1} \left[1-G_{j} \right] \kh^{2} d\kh d\mu} 
\end{equation}

where $G_{c}=2\pi \gk^{-1}$ with $\vzk=v_{c\k}$ and $G_{ab}$ is given by:

\begin{equation} \label{Gab}
G_{ab} = \frac{2\pi}{ \sqrt{(1 + i \eta \tk)^{2} + x_{ab}^{2}}}
\end{equation}
Equation \ref{Gab} was obtained through the use of identity 3.661.3 in Ref.\ \onlinecite{Rzhik} to perform the integral over $\phi$.

The heat flux $\widetilde{q_{j}}$ is given by:
\begin{eqnarray} \label{}
q_{j} &=& A \int_{-1}^{1} \int_{0}^{1} g_{\k} v_{j\k} H_{j} \kh^{2} d\kh d\mu  
\end{eqnarray}

where $H_{c}=2\pi \gk^{-1}$ and 
\begin{equation} \label{Hab}
H_{ab}=\frac{2\pi }{ix_{ab}} \left(1-\sqrt{ \frac{(1+i \eta \tk)^{2} }{(1+i \eta \tk)^{2} + x_{ab}^{2} } } \right)
\end{equation}

This result was obtained using the following identity:

\begin{equation} \label{}
\int_{0}^{2\pi} \frac{\cos \phi}{a + b \cos \phi} d\phi = \frac{2\pi}{b} \left( 1 - \sqrt{ \frac{a^{2}}{a^{2} - b^{2}} } \right)
\end{equation}

The thermal conductivity $\kappa_{j}$ can be written:
\begin{equation} \label{}
\kappa_{j} = A_{j} \int_{-1}^{1} \int_{0}^{1} C_{\k} v_{j,k}^{2} \tk \kh^{2} d\kh d\mu = A_{j} \int_{-1}^{1} \int_{0}^{1} \kappa_{j\k}  \kh^{2} d\kh d\mu 
\end{equation}

where $A_{c} = 2\pi A$ and $A_{ab} = \pi A$. Finally, the effective thermal conductivity in the weakly quasiballistic regime presented in Sec.\ \ref{sec:wqb} is given by:

\begin{equation} \label{}
\kappa_{eff,j} = 2\pi A \int_{-1}^{1} \int_{0}^{1} \kappa_{j\k} S_{j} \kh^{2} d\kh d\mu
\end{equation}

where $S_{c}=(1+x_{c}^{-2})^{-1}$ and 

\begin{equation} \label{}
S_{ab} = \frac{1}{x_{ab}^{2}} \left( 1 - \frac{1}{\sqrt{1+x_{ab}^{2}}} \right)
\end{equation}
where $x_{j}=\xi_{z} \Lambda_{j\k}$. This result was obtained using the relation:

\begin{equation} \label{}
\int_{0}^{2\pi} \frac{\cos^{2} \phi}{a + b \cos^{2} \phi} d\phi = \frac{2\pi}{b} \left( 1 - \sqrt{ \frac{a}{a+b} } \right)
\end{equation}

To model graphite, we take the sound velocities to be $v_{ab}=15$ km/s and $v_{c}=1$ km/s. The relationship between the number density and the maximum wavevectors to ensure the correct number of modes is \cite{chen_anisotropic_2013}
\begin{equation} \label{}
n_{puc} = \frac{1}{6 \pi^{2}} (k_{c,m} k_{ab,m}^{2}) ^{1/3}
\end{equation}
We consider only acoustic phonons and thus $n_{puc}=n/4$ represents the number density of primitive unit cells, where $n=1.13\times10^{29}$ m$^{-3}$ is the atomic number density in graphite. Incorporating the anisotropy of the actual lattice constant of graphite, we obtain $k_{ab,m}=13$ nm$^{-1}$ and $k_{c,m}=9.7 $ nm$^{-1}$. These maximum wavevectors correspond to lattice constants of $a_{ab}=2.38$ $\AA$ and $a_{c}=3.32$ $\AA$, quite close to the actual lattice constants of $a_{ab}=2.46$ $\AA$ and $a_{c}=3.35$ $\AA$, respectively. The relaxation times are chosen to be:

\begin{equation} \label{}
\tau^{-1} = \frac{10^{17}}{\omega(\k)^{2}} + \frac{v}{L_{b}}
\end{equation}
where $L_{b}=1$ mm is a boundary scattering length scale and $v$ is the velocity of a particular mode. Note that we assume the relaxation time to be isotropic. We assume that 3 identical polarizations exist. These parameters yield $\kappa_{ab}=2680$ W/mK and $\kappa_{c}=4$ W/mK, reasonably close to the literature values of 2000 W/mK and 6 W/mK, respectively. \cite{touloukian_thermophysical_1970}

\section{Results} \label{sec:results}

\subsection{Thermal conductivity}

We first investigate the correlation between the thermal conductivity along the c- and ab-axes by calculating the thermal conductivities as a function of c- and ab-axis group velocity. While similar equations as Eq.\ \ref{eq:kz} for the thermal conductivity of graphite have been reported by Slack \cite{slack_anisotropic_1962} and Hsieh et al. \cite{hsieh_pressure_2009}, to our knowledge no calculation of thermal conductivity versus group velocity along different crystallographic axes has been reported. The closest result is the recent report by Wei et al. of thermal conductivity versus a bonding strength parameter in MD simulations \cite{wei_negative_2013}.

\begin{figure}
\begin{center}
\includegraphics[width=1\textwidth]{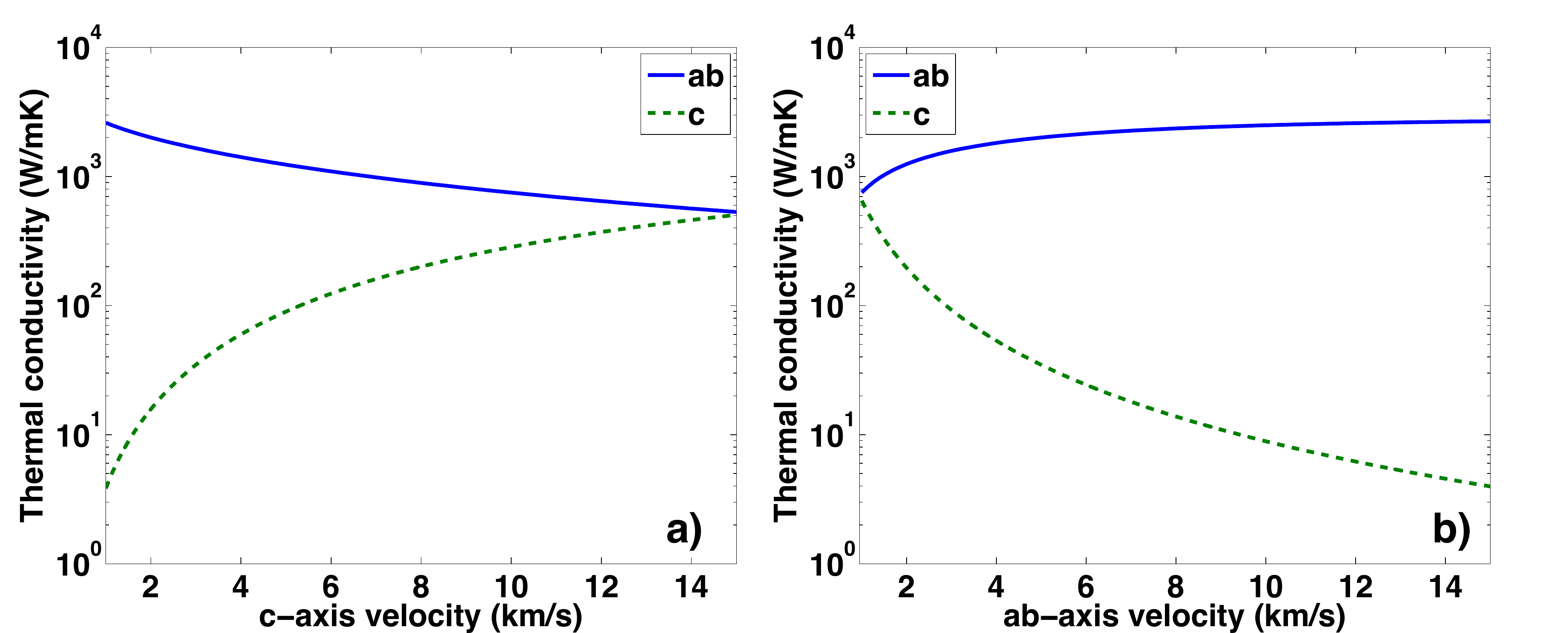}
\caption{Thermal conductivity of graphite along the ab-axis and c-axis as the (a) c-axis velocity is varied and ab-axis velocity is fixed at 15 km/s, and (b) ab-axis velocity is varied and the c-axis velocity is fixed at 1 km/s. Decreasing the c-axis velocity increases the ab-axis thermal conductivity while decreasing the ab-axis velocity increases the c-axis thermal conductivity. }
\label{fig:kvel}
\end{center}
\end{figure}

Figure \ref{fig:kvel}a and b plots the ab- and c-axis thermal conductivities as a function of the c-axis or ab-axis group velocity, respectively. This plot shows the expected result  that decreasing the group velocity also decreases the thermal conductivity along that direction. However, the plot also reveals the counterintuitive result that the thermal conductivity increases along the perpendicular direction as this group velocity decreases. For example, in Fig.\ \ref{fig:kvel}b, decreasing the ab-axis velocity increases the c-axis thermal conductivity.

The reason for this counterintuitive trend has been described in detail by Wei et al. \cite{wei_negative_2013} and is due to the highly elliptical isofrequency surface. The group velocity along a particular crystallographic direction equals the component of the gradient of the isofrequency surface along that direction. As the fast direction group velocity decreases the Brillouin zone becomes more spherical and increases the average group velocity along the slow direction, leading to a higher thermal conductivity along the slow direction even while the thermal conductivity decreases in the fast direction. While Ref.\ \onlinecite{wei_negative_2013} used MD simulations and varied bonding strengths to identify this negative correlation, our calculation shows that the same trend can be observed simply by incorporating anisotropic group velocities into the BTE. This result demonstrates that the shape of the isofrequency surface, as determined by the dispersion relation, plays a critical role in setting the thermal conductivity along all crystallographic directions.

We point out another counterintuitive result in the isotropic limit: the thermal conductivity when the group velocity along all three directions is 1 km/s, Fig.\ \ref{fig:kvel}b, is slightly larger than that when all group velocities equal 15 km/s, Fig.\ \ref{fig:kvel}a, though from the kinetic equation the opposite trend is expected. The discrepancy can be explained by differences in the phonon frequency spectrum in each direction. In the case where the group velocity is 1 km/s, the maximum frequency is around 2 THz and all modes are occupied since $k_{B}T/h \approx 6$ THz at $T=300$ K. In the latter case, the maximum phonon frequency is around 31 THz and many modes remain unoccupied, resulting in a smaller specific heat and slightly smaller thermal conductivity than in the former case. Thus when considering the origin of thermal conductivity anisotropy it is important to compare the occupation of modes along different directions. The difference in occupation is particularly large for graphite because many of the stiff, high frequency in-plane modes remain unoccupied at room temperature while essentially all of the soft, low frequency cross-plane modes are occupied.

Also, note from Fig.\ \ref{fig:kvel}b that the thermal conductivity when all group velocities equal 1 km/s is orders of magnitude larger than the c-axis value of graphite. This observation highlights that a primary reason that the c-axis thermal conductivity is low is actually because the ab-axis group velocity, and by extension the ab-axis thermal conductivity, is high. The difference in group velocities lead to a highly elliptical isofrequency surface that decreases the group velocity along the c-axis as discussed above and in Ref.  \onlinecite{wei_negative_2013}.

\subsection{Quasiballistic regime and MFP reconstruction}

\begin{figure}
\begin{center}
\includegraphics[width=1\textwidth]{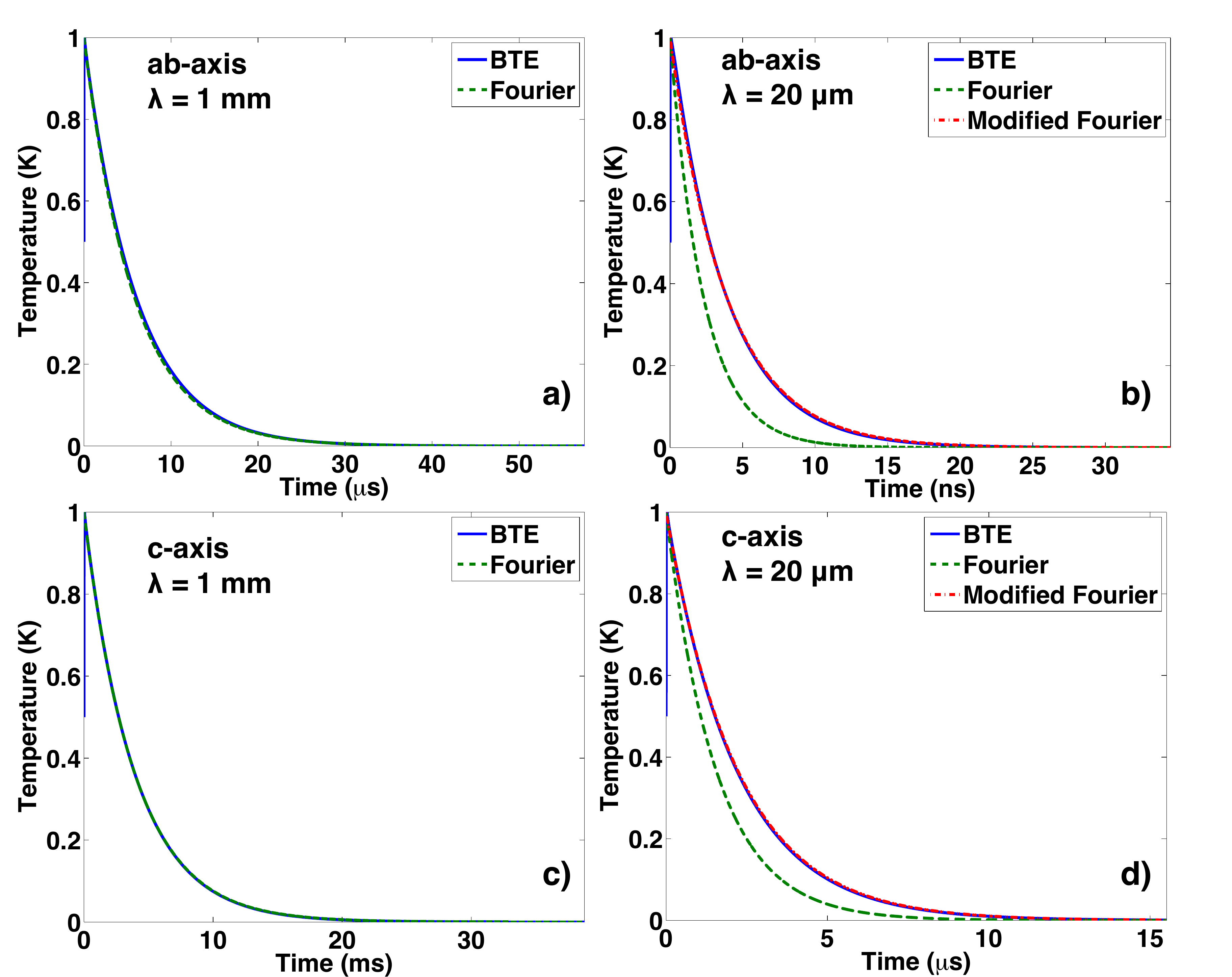}
\caption{Temperature decay versus time for a spatially sinusoidal heat impulse with grating period $\lambda$ for the (a) ab-axis and $\lambda= 1$ mm; (b) ab-axis and $\lambda= 20$ $\mu$m; (c) c-axis and $\lambda= 1$ mm; (d) c-axis and $\lambda= 20$ $\mu$m. In the diffusive limit, (a) and (c), the BTE solution agrees with diffusion theory. In the quasiballistic limit, (b) and (d), the BTE solution agrees with a modified Fourier law with a thermal conductivity given by Eq.\ \ref{eq:keff}. Note that quasiballistic effects are apparent along both the ab- and c-axes even though the thermal conductivity is much smaller along the c-axis.}
\label{Tdecay}
\end{center}
\end{figure}

We now use our solution to investigate the dynamics of thermal transport in the quasiballistic regime in which thermal gradients occur over length scales comparable to MFPs. First, we calculate the transient temperature response assuming the heating profile is a spatially sinusoidal impulse with period $\lambda = 2 \pi / \xi_{x}$ as occurs in transient grating experiments \cite{johnson_direct_2013}. The result is obtained by performing an inverse fast Fourier transform on Eq.\ \ref{DTgraphite} and is shown in Fig.\ \ref{Tdecay}. For a grating period of 1 mm along the ab-axis, the heat conduction is diffusive and the BTE solution agrees with Fourier's law, as in Fig.\ \ref{Tdecay}a. For a grating period of 20 $\mu$m, Fig.\ \ref{Tdecay}b, the thermal decay deviates from the Fourier's law prediction, indicating that MFPs are comparable to the grating period. Such long MFPs could be expected given the large thermal conductivity along the ab-axis. The thermal decay from the BTE is in excellent agreement with a modified Fourier's law with a thermal conductivity given by Eq.\ \ref{eq:keff}, as expected based on the derivation in Sec.\ \ref{sec:wqb}.

The corresponding results for the c-axis are presented in Figs.\ \ref{Tdecay}c and \ref{Tdecay}d. Similarly to the ab-axis, the heat conduction is diffusive for a grating period of 1 mm. Unexpectedly, however, significant quasiballistic effects are observed for a grating period of 20 $\mu$m in Fig.\ \ref{Tdecay}d. This observation is counterintuitive because the c-axis thermal conductivity is 250 times smaller than the ab-axis value, yet the c-axis MFPs appear to comparable to those along the ab-axis. The apparently long c-axis phonon MFPs were previously observed using MD simulations.\cite{wei_phonon_2014}

In fact, long c-axis MFP phonons should not be entirely unexpected despite the low c-axis thermal conductivity. Recall that we have assumed the relaxation times to depend only on the phonon frequency. Further, only low frequency phonons are supported along the c-axis, and these phonons have long relaxation times and hence long MFPs. The c-axis MFPs thus are only smaller than those along the ab-axis by a factor on the order of the velocity anisotropy ratio equal to 15 for graphite. However, due to the low phonon frequencies supported in the cross-plane direction these long MFPs contribute substantially to c-axis thermal conductivity as higher frequency phonons with shorter MFPs do not exist along the c-axis. In this way, the c-axis thermal conductivity can be small even though the phonons carrying the heat can have long MFPs. This result highlights the importance of considering the different phonon frequencies that are supported along each direction when analyzing anisotropic crystals.

To gain more insight into the MFPs responsible for heat conduction along each crystallographic axis, we now investigate whether the MFP spectrum can be extracted from the transient thermal decays in Fig.\ \ref{Tdecay}. Ideally this extraction could be performed solely from the quasiballistic observations without requiring any prior knowledge of the harmonic properties or anisotropy of the lattice. Minnich presented a reconstruction approach in which the MFP spectrum of an isotropic solid can be obtained from thermal conductivity measurements in the quasiballistic regime \cite{minnich_determining_2012}, but this approach assumed isotropy and cannot be applied to arbitrary solids in its present form.

To see how to recover the anisotropic MFP spectrum, consider Eq.\ \ref{eq:keff}, which expresses the effective thermal conductivity $\kappa_{eff}$ as a function of the spectral thermal conductivity $\kappa_{z\k}$ and a suppression function $S(x)$, where $x=\xi_{z} \lk$. This equation has exactly the same form as that presented in Ref.\ \onlinecite{minnich_determining_2012} with the exception that the MFP is no longer the isotropic MFP $\Lambda$ but rather the component of the MFP along the transport direction $\lk$. This observation suggests that a generalized reconstruction approach can be implemented to obtain the MFP spectrum of a crystal with an arbitrary dispersion relation by letting the independent variable be $\lk$ rather than $\Lambda$. The sum over phase space can then transformed into an integral over $\lk$, yielding an equation exactly analogous to that considered by Chen and Dames \cite{chen_anisotropic_2013} and Minnich \cite{minnich_determining_2012}. Then, the anisotropic MFP spectrum can be obtained from the effective thermal conductivities using the method of Ref.\ \onlinecite{minnich_determining_2012} without any knowledge of the anisotropy, harmonic properties, or the MFPs of the material of interest.

\begin{figure}
\begin{center}
\includegraphics[width=1\textwidth]{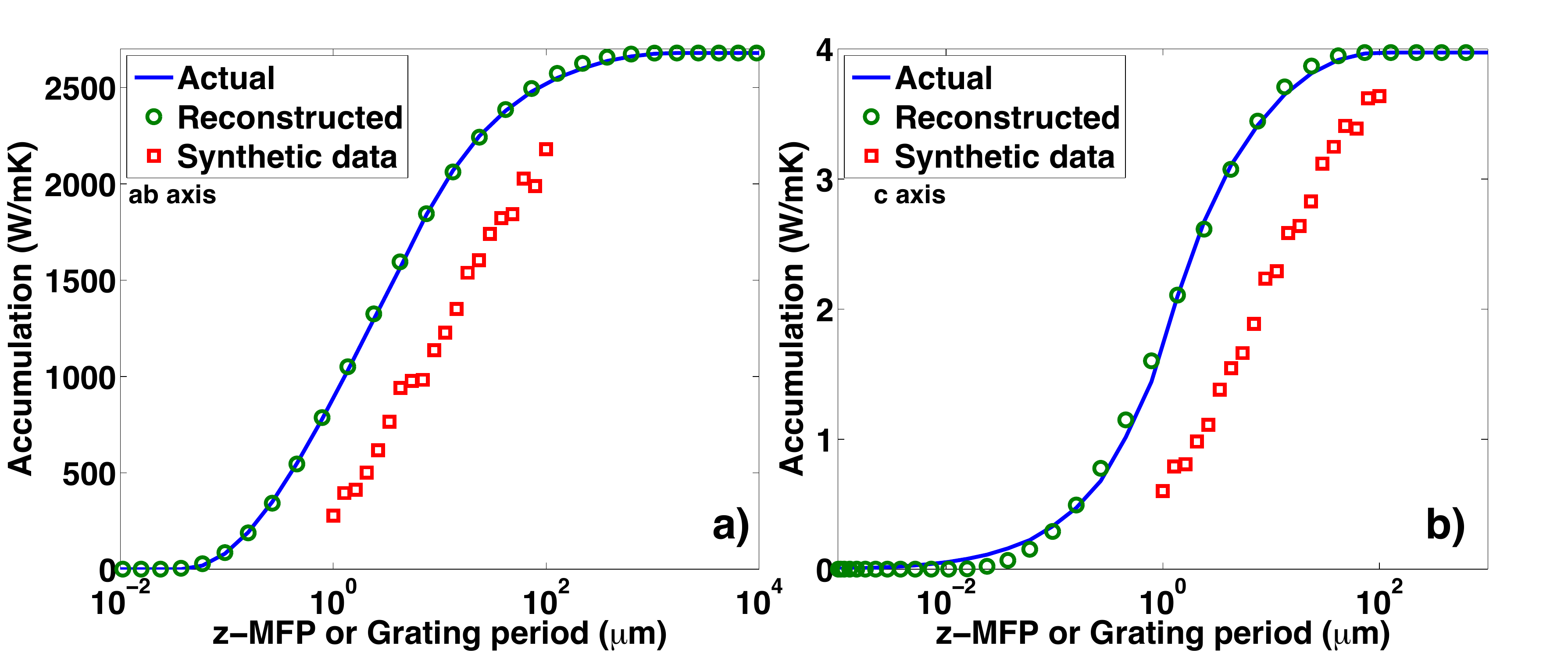}
\caption{Accumulated MFP spectrum versus component of MFP along the transport direction (z-MFP) or grating period (for synthetic data) along the (a) ab-axis and (b) c-axis. The synthesized thermal conductivities with added random noise are the red squares, the actual MFP spectrum is the blue line, and the result of the reconstruction are the green circles. Our reconstruction approach can determine the generalized MFP spectrum without any knowledge of the harmonic or anharmonic properties of the crystal.}
\label{MFPr}
\end{center}
\end{figure}

We demonstrate this approach using the numerical method exactly as described in Ref.\ \onlinecite{minnich_determining_2012}. Briefly, we synthesize effective thermal conductivities numerically using Eq.\ \ref{eq:keff} and add artificial random noise to simulate experimental uncertainty. Using these effective thermal conductivities and our knowledge of the suppression function, we use CVX \cite{gb08,cvx} to solve for the generalized MFP spectrum. The result is shown in Fig.\ \ref{MFPr}, demonstrating excellent agreement between the reconstructed and actual MFP spectra along both the ab- and c-axes. This result shows that our approach can reconstruct the anisotropic MFP spectrum along different crystallographic directions with only the effective thermal conductivities and the known suppression function as input. 


The MFP spectra in Fig.\ \ref{MFPr} demonstrate that the MFPs responsible for heat conduction along the ab- and c-axes can be comparable even though the calculated thermal conductivities differ by a factor of over 600. This result indicates that in anisotropic crystals scattering is not the sole origin of low thermal conductivity as is typically assumed for isotropic crystals. Instead, low thermal conductivity can originate from the group velocity and phonon frequencies that are supported along each crystallographic direction as determined by the shape of the isofrequency surface. Clearly, properly incorporating a realistic shape of the isofrequency surface is essential to making accurate predictions of thermal transport in anisotropic crystals. In the present case, isotropic approximations will yield predictions that are qualitatively incorrect.

Finally, we note that the MFPs in actual graphite are unlikely to be as long as those presented in Fig.\ \ref{MFPr}. MD simulations report the c-axis MFPs to be on the order of a few hundred nanometers \cite{wei_phonon_2014}. The predictions in the present work could likely be improved by incorporating the exact dispersion for graphite from lattice dynamics, an effort that we will address in a future work.

\section{Summary}

We have used an analytic solution of the BTE to study heat conduction in graphite. We find that the thermal conductivity anisotropy of these crystals can be mostly explained by the anisotropic group velocity, and that the MFPs of the thermal phonons can be comparable along each direction even while the thermal conductivities differ substantially. Further, we have described a generalized MFP spectrum reconstruction approach that enables the anisotropic MFP spectrum to be reconstructed without any prior knowledge of the properties of the crystal. Our result provides a better understanding of heat conduction in anisotropic crystals and the theoretical basis necessary to directly measure the phonon MFP spectrum in a wide range of solids.

\section{Acknowledgements}

The author thanks Chengyun Hua and Ding Ding for performing several integrals. This work was sponsored in part by Robert Bosch LLC through Bosch Energy Research Network Grant no. 13.01.CC11, by the National Science Foundation under Grant no. CBET CAREER 1254213, and by Boeing under the Boeing-Caltech Strategic Research \& Development Relationship Agreement.

\bibliography{Anisotropy}

\begin{thebibliography}{44}
\expandafter\ifx\csname natexlab\endcsname\relax\def\natexlab#1{#1}\fi
\expandafter\ifx\csname bibnamefont\endcsname\relax
  \def\bibnamefont#1{#1}\fi
\expandafter\ifx\csname bibfnamefont\endcsname\relax
  \def\bibfnamefont#1{#1}\fi
\expandafter\ifx\csname citenamefont\endcsname\relax
  \def\citenamefont#1{#1}\fi
\expandafter\ifx\csname url\endcsname\relax
  \def\url#1{\texttt{#1}}\fi
\expandafter\ifx\csname urlprefix\endcsname\relax\def\urlprefix{URL }\fi
\providecommand{\bibinfo}[2]{#2}
\providecommand{\eprint}[2][]{\url{#2}}

\bibitem[{\citenamefont{Touloukian}(1970)}]{touloukian_thermophysical_1970}
\bibinfo{author}{\bibfnamefont{Y.~S.} \bibnamefont{Touloukian}},
  \emph{\bibinfo{title}{Thermophysical Properties of Matter: Thermal
  conductivity : nonmetallic solids}} (\bibinfo{publisher}{{IFI}/Plenum},
  \bibinfo{year}{1970}), ISBN \bibinfo{isbn}{9780306670206}.

\bibitem[{\citenamefont{Jang et~al.}(2010)\citenamefont{Jang, Chen, Bao, Lau,
  and Dames}}]{jang_thickness-dependent_2010}
\bibinfo{author}{\bibfnamefont{W.}~\bibnamefont{Jang}},
  \bibinfo{author}{\bibfnamefont{Z.}~\bibnamefont{Chen}},
  \bibinfo{author}{\bibfnamefont{W.}~\bibnamefont{Bao}},
  \bibinfo{author}{\bibfnamefont{C.~N.} \bibnamefont{Lau}}, \bibnamefont{and}
  \bibinfo{author}{\bibfnamefont{C.}~\bibnamefont{Dames}},
  \bibinfo{journal}{Nano Letters} \textbf{\bibinfo{volume}{10}},
  \bibinfo{pages}{3909} (\bibinfo{year}{2010}), ISSN \bibinfo{issn}{1530-6984},
  \urlprefix\url{http://dx.doi.org/10.1021/nl101613u}.

\bibitem[{\citenamefont{Ghosh et~al.}(2010)\citenamefont{Ghosh, Bao, Nika,
  Subrina, Pokatilov, Lau, and Balandin}}]{ghosh_dimensional_2010}
\bibinfo{author}{\bibfnamefont{S.}~\bibnamefont{Ghosh}},
  \bibinfo{author}{\bibfnamefont{W.}~\bibnamefont{Bao}},
  \bibinfo{author}{\bibfnamefont{D.~L.} \bibnamefont{Nika}},
  \bibinfo{author}{\bibfnamefont{S.}~\bibnamefont{Subrina}},
  \bibinfo{author}{\bibfnamefont{E.~P.} \bibnamefont{Pokatilov}},
  \bibinfo{author}{\bibfnamefont{C.~N.} \bibnamefont{Lau}}, \bibnamefont{and}
  \bibinfo{author}{\bibfnamefont{A.~A.} \bibnamefont{Balandin}},
  \bibinfo{journal}{Nature Materials} \textbf{\bibinfo{volume}{9}},
  \bibinfo{pages}{555} (\bibinfo{year}{2010}), ISSN \bibinfo{issn}{1476-1122},
  \urlprefix\url{http://www.nature.com/nmat/journal/v9/n7/full/nmat2753.html}.

\bibitem[{\citenamefont{Seol et~al.}(2010)\citenamefont{Seol, Jo, Moore,
  Lindsay, Aitken, Pettes, Li, Yao, Huang, Broido
  et~al.}}]{seol_two-dimensional_2010}
\bibinfo{author}{\bibfnamefont{J.~H.} \bibnamefont{Seol}},
  \bibinfo{author}{\bibfnamefont{I.}~\bibnamefont{Jo}},
  \bibinfo{author}{\bibfnamefont{A.~L.} \bibnamefont{Moore}},
  \bibinfo{author}{\bibfnamefont{L.}~\bibnamefont{Lindsay}},
  \bibinfo{author}{\bibfnamefont{Z.~H.} \bibnamefont{Aitken}},
  \bibinfo{author}{\bibfnamefont{M.~T.} \bibnamefont{Pettes}},
  \bibinfo{author}{\bibfnamefont{X.}~\bibnamefont{Li}},
  \bibinfo{author}{\bibfnamefont{Z.}~\bibnamefont{Yao}},
  \bibinfo{author}{\bibfnamefont{R.}~\bibnamefont{Huang}},
  \bibinfo{author}{\bibfnamefont{D.}~\bibnamefont{Broido}},
  \bibnamefont{et~al.}, \bibinfo{journal}{Science}
  \textbf{\bibinfo{volume}{328}}, \bibinfo{pages}{213} (\bibinfo{year}{2010}),
  ISSN \bibinfo{issn}{0036-8075, 1095-9203},
  \urlprefix\url{http://www.sciencemag.org/content/328/5975/213}.

\bibitem[{\citenamefont{Bonini et~al.}(2012)\citenamefont{Bonini, Garg, and
  Marzari}}]{bonini_acoustic_2012}
\bibinfo{author}{\bibfnamefont{N.}~\bibnamefont{Bonini}},
  \bibinfo{author}{\bibfnamefont{J.}~\bibnamefont{Garg}}, \bibnamefont{and}
  \bibinfo{author}{\bibfnamefont{N.}~\bibnamefont{Marzari}},
  \bibinfo{journal}{Nano Letters} \textbf{\bibinfo{volume}{12}},
  \bibinfo{pages}{2673} (\bibinfo{year}{2012}), ISSN \bibinfo{issn}{1530-6984},
  \urlprefix\url{http://dx.doi.org/10.1021/nl202694m}.

\bibitem[{\citenamefont{Lindsay et~al.}(2014)\citenamefont{Lindsay, Li,
  Carrete, Mingo, Broido, and Reinecke}}]{lindsay_phonon_2014}
\bibinfo{author}{\bibfnamefont{L.}~\bibnamefont{Lindsay}},
  \bibinfo{author}{\bibfnamefont{W.}~\bibnamefont{Li}},
  \bibinfo{author}{\bibfnamefont{J.}~\bibnamefont{Carrete}},
  \bibinfo{author}{\bibfnamefont{N.}~\bibnamefont{Mingo}},
  \bibinfo{author}{\bibfnamefont{D.~A.} \bibnamefont{Broido}},
  \bibnamefont{and} \bibinfo{author}{\bibfnamefont{T.~L.}
  \bibnamefont{Reinecke}}, \bibinfo{journal}{Physical Review B}
  \textbf{\bibinfo{volume}{89}}, \bibinfo{pages}{155426}
  (\bibinfo{year}{2014}),
  \urlprefix\url{http://link.aps.org/doi/10.1103/PhysRevB.89.155426}.

\bibitem[{\citenamefont{Paulatto et~al.}(2013)\citenamefont{Paulatto, Mauri,
  and Lazzeri}}]{paulatto_anharmonic_2013}
\bibinfo{author}{\bibfnamefont{L.}~\bibnamefont{Paulatto}},
  \bibinfo{author}{\bibfnamefont{F.}~\bibnamefont{Mauri}}, \bibnamefont{and}
  \bibinfo{author}{\bibfnamefont{M.}~\bibnamefont{Lazzeri}},
  \bibinfo{journal}{Physical Review B} \textbf{\bibinfo{volume}{87}},
  \bibinfo{pages}{214303} (\bibinfo{year}{2013}),
  \urlprefix\url{http://link.aps.org/doi/10.1103/PhysRevB.87.214303}.

\bibitem[{\citenamefont{Bae et~al.}(2013)\citenamefont{Bae, Li, Aksamija,
  Martin, Xiong, Ong, Knezevic, and Pop}}]{bae_ballistic_2013}
\bibinfo{author}{\bibfnamefont{M.-H.} \bibnamefont{Bae}},
  \bibinfo{author}{\bibfnamefont{Z.}~\bibnamefont{Li}},
  \bibinfo{author}{\bibfnamefont{Z.}~\bibnamefont{Aksamija}},
  \bibinfo{author}{\bibfnamefont{P.~N.} \bibnamefont{Martin}},
  \bibinfo{author}{\bibfnamefont{F.}~\bibnamefont{Xiong}},
  \bibinfo{author}{\bibfnamefont{Z.-Y.} \bibnamefont{Ong}},
  \bibinfo{author}{\bibfnamefont{I.}~\bibnamefont{Knezevic}}, \bibnamefont{and}
  \bibinfo{author}{\bibfnamefont{E.}~\bibnamefont{Pop}},
  \bibinfo{journal}{Nature Communications} \textbf{\bibinfo{volume}{4}},
  \bibinfo{pages}{1734} (\bibinfo{year}{2013}),
  \urlprefix\url{http://www.nature.com/ncomms/journal/v4/n4/full/ncomms2755.html}.

\bibitem[{\citenamefont{Xu et~al.}(2014)\citenamefont{Xu, Pereira, Wang, Wu,
  Zhang, Zhao, Bae, Tinh~Bui, Xie, Thong et~al.}}]{xu_length-dependent_2014}
\bibinfo{author}{\bibfnamefont{X.}~\bibnamefont{Xu}},
  \bibinfo{author}{\bibfnamefont{L.~F.~C.} \bibnamefont{Pereira}},
  \bibinfo{author}{\bibfnamefont{Y.}~\bibnamefont{Wang}},
  \bibinfo{author}{\bibfnamefont{J.}~\bibnamefont{Wu}},
  \bibinfo{author}{\bibfnamefont{K.}~\bibnamefont{Zhang}},
  \bibinfo{author}{\bibfnamefont{X.}~\bibnamefont{Zhao}},
  \bibinfo{author}{\bibfnamefont{S.}~\bibnamefont{Bae}},
  \bibinfo{author}{\bibfnamefont{C.}~\bibnamefont{Tinh~Bui}},
  \bibinfo{author}{\bibfnamefont{R.}~\bibnamefont{Xie}},
  \bibinfo{author}{\bibfnamefont{J.~T.~L.} \bibnamefont{Thong}},
  \bibnamefont{et~al.}, \bibinfo{journal}{Nature Communications}
  \textbf{\bibinfo{volume}{5}} (\bibinfo{year}{2014}),
  \urlprefix\url{http://www.nature.com/ncomms/2014/140416/ncomms4689/full/ncomms4689.html}.

\bibitem[{\citenamefont{Yan et~al.}(2012)\citenamefont{Yan, Liu, Khan, and
  Balandin}}]{yan_graphene_2012}
\bibinfo{author}{\bibfnamefont{Z.}~\bibnamefont{Yan}},
  \bibinfo{author}{\bibfnamefont{G.}~\bibnamefont{Liu}},
  \bibinfo{author}{\bibfnamefont{J.~M.} \bibnamefont{Khan}}, \bibnamefont{and}
  \bibinfo{author}{\bibfnamefont{A.~A.} \bibnamefont{Balandin}},
  \bibinfo{journal}{Nature Communications} \textbf{\bibinfo{volume}{3}},
  \bibinfo{pages}{827} (\bibinfo{year}{2012}), ISSN \bibinfo{issn}{2041-1723},
  \urlprefix\url{http://www.nature.com/ncomms/journal/v3/n5/full/ncomms1828.html}.

\bibitem[{\citenamefont{Chiritescu et~al.}(2007)\citenamefont{Chiritescu,
  Cahill, Nguyen, Johnson, Bodapati, Keblinski, and
  Zschack}}]{chiritescu_ultralow_2007}
\bibinfo{author}{\bibfnamefont{C.}~\bibnamefont{Chiritescu}},
  \bibinfo{author}{\bibfnamefont{D.~G.} \bibnamefont{Cahill}},
  \bibinfo{author}{\bibfnamefont{N.}~\bibnamefont{Nguyen}},
  \bibinfo{author}{\bibfnamefont{D.}~\bibnamefont{Johnson}},
  \bibinfo{author}{\bibfnamefont{A.}~\bibnamefont{Bodapati}},
  \bibinfo{author}{\bibfnamefont{P.}~\bibnamefont{Keblinski}},
  \bibnamefont{and} \bibinfo{author}{\bibfnamefont{P.}~\bibnamefont{Zschack}},
  \bibinfo{journal}{Science} \textbf{\bibinfo{volume}{315}},
  \bibinfo{pages}{351} (\bibinfo{year}{2007}), ISSN \bibinfo{issn}{0036-8075,
  1095-9203}, \urlprefix\url{http://www.sciencemag.org/content/315/5810/351}.

\bibitem[{\citenamefont{Losego et~al.}(2013)\citenamefont{Losego, Blitz, Vaia,
  Cahill, and Braun}}]{losego_ultralow_2013}
\bibinfo{author}{\bibfnamefont{M.~D.} \bibnamefont{Losego}},
  \bibinfo{author}{\bibfnamefont{I.~P.} \bibnamefont{Blitz}},
  \bibinfo{author}{\bibfnamefont{R.~A.} \bibnamefont{Vaia}},
  \bibinfo{author}{\bibfnamefont{D.~G.} \bibnamefont{Cahill}},
  \bibnamefont{and} \bibinfo{author}{\bibfnamefont{P.~V.} \bibnamefont{Braun}},
  \bibinfo{journal}{Nano Letters}  (\bibinfo{year}{2013}), ISSN
  \bibinfo{issn}{1530-6984},
  \urlprefix\url{http://dx.doi.org/10.1021/nl4007326}.

\bibitem[{\citenamefont{Zhao et~al.}(2014)\citenamefont{Zhao, Lo, Zhang, Sun,
  Tan, Uher, Wolverton, Dravid, and Kanatzidis}}]{zhao_ultralow_2014}
\bibinfo{author}{\bibfnamefont{L.-D.} \bibnamefont{Zhao}},
  \bibinfo{author}{\bibfnamefont{S.-H.} \bibnamefont{Lo}},
  \bibinfo{author}{\bibfnamefont{Y.}~\bibnamefont{Zhang}},
  \bibinfo{author}{\bibfnamefont{H.}~\bibnamefont{Sun}},
  \bibinfo{author}{\bibfnamefont{G.}~\bibnamefont{Tan}},
  \bibinfo{author}{\bibfnamefont{C.}~\bibnamefont{Uher}},
  \bibinfo{author}{\bibfnamefont{C.}~\bibnamefont{Wolverton}},
  \bibinfo{author}{\bibfnamefont{V.~P.} \bibnamefont{Dravid}},
  \bibnamefont{and} \bibinfo{author}{\bibfnamefont{M.~G.}
  \bibnamefont{Kanatzidis}}, \bibinfo{journal}{Nature}
  \textbf{\bibinfo{volume}{508}}, \bibinfo{pages}{373} (\bibinfo{year}{2014}),
  ISSN \bibinfo{issn}{0028-0836},
  \urlprefix\url{http://www.nature.com.clsproxy.library.caltech.edu/nature/journal/v508/n7496/full/nature13184.html}.

\bibitem[{\citenamefont{von Gutfeld and
  Nethercot}(1964)}]{von_gutfeld_heat_1964}
\bibinfo{author}{\bibfnamefont{R.~J.} \bibnamefont{von Gutfeld}}
  \bibnamefont{and} \bibinfo{author}{\bibfnamefont{A.~H.}
  \bibnamefont{Nethercot}}, \bibinfo{journal}{Physical Review Letters}
  \textbf{\bibinfo{volume}{12}}, \bibinfo{pages}{641} (\bibinfo{year}{1964}),
  \urlprefix\url{http://link.aps.org/doi/10.1103/PhysRevLett.12.641}.

\bibitem[{\citenamefont{Al-Jishi and
  Dresselhaus}(1982)}]{al-jishi_lattice-dynamical_1982}
\bibinfo{author}{\bibfnamefont{R.}~\bibnamefont{Al-Jishi}} \bibnamefont{and}
  \bibinfo{author}{\bibfnamefont{G.}~\bibnamefont{Dresselhaus}},
  \bibinfo{journal}{Physical Review B} \textbf{\bibinfo{volume}{26}},
  \bibinfo{pages}{4514} (\bibinfo{year}{1982}),
  \urlprefix\url{http://link.aps.org/doi/10.1103/PhysRevB.26.4514}.

\bibitem[{\citenamefont{Nicklow et~al.}(1972)\citenamefont{Nicklow,
  Wakabayashi, and Smith}}]{nicklow_lattice_1972}
\bibinfo{author}{\bibfnamefont{R.}~\bibnamefont{Nicklow}},
  \bibinfo{author}{\bibfnamefont{N.}~\bibnamefont{Wakabayashi}},
  \bibnamefont{and} \bibinfo{author}{\bibfnamefont{H.~G.} \bibnamefont{Smith}},
  \bibinfo{journal}{Physical Review B} \textbf{\bibinfo{volume}{5}},
  \bibinfo{pages}{4951} (\bibinfo{year}{1972}),
  \urlprefix\url{http://link.aps.org/doi/10.1103/PhysRevB.5.4951}.

\bibitem[{\citenamefont{Bowman and
  Krumhansl}(1958)}]{bowman_low-temperature_1958}
\bibinfo{author}{\bibfnamefont{J.~C.} \bibnamefont{Bowman}} \bibnamefont{and}
  \bibinfo{author}{\bibfnamefont{J.~A.} \bibnamefont{Krumhansl}},
  \bibinfo{journal}{Journal of Physics and Chemistry of Solids}
  \textbf{\bibinfo{volume}{6}}, \bibinfo{pages}{367} (\bibinfo{year}{1958}),
  ISSN \bibinfo{issn}{0022-3697},
  \urlprefix\url{http://www.sciencedirect.com/science/article/pii/0022369758900556}.

\bibitem[{\citenamefont{Krumhansl and Brooks}(1953)}]{krumhansl_lattice_1953}
\bibinfo{author}{\bibfnamefont{J.}~\bibnamefont{Krumhansl}} \bibnamefont{and}
  \bibinfo{author}{\bibfnamefont{H.}~\bibnamefont{Brooks}},
  \bibinfo{journal}{The Journal of Chemical Physics}
  \textbf{\bibinfo{volume}{21}}, \bibinfo{pages}{1663} (\bibinfo{year}{1953}),
  ISSN \bibinfo{issn}{0021-9606, 1089-7690},
  \urlprefix\url{http://scitation.aip.org/content/aip/journal/jcp/21/10/10.1063/1.1698641}.

\bibitem[{\citenamefont{Slack}(1962)}]{slack_anisotropic_1962}
\bibinfo{author}{\bibfnamefont{G.~A.} \bibnamefont{Slack}},
  \bibinfo{journal}{Physical Review} \textbf{\bibinfo{volume}{127}},
  \bibinfo{pages}{694} (\bibinfo{year}{1962}),
  \urlprefix\url{http://link.aps.org/doi/10.1103/PhysRev.127.694}.

\bibitem[{\citenamefont{Taylor}(1966)}]{taylor_thermal_1966}
\bibinfo{author}{\bibfnamefont{R.}~\bibnamefont{Taylor}},
  \bibinfo{journal}{Philosophical Magazine} \textbf{\bibinfo{volume}{13}},
  \bibinfo{pages}{157} (\bibinfo{year}{1966}), ISSN \bibinfo{issn}{0031-8086},
  \urlprefix\url{http://dx.doi.org/10.1080/14786436608211993}.

\bibitem[{\citenamefont{Tanaka and Suzuki}(1972)}]{tanaka_thermal_1972}
\bibinfo{author}{\bibfnamefont{T.}~\bibnamefont{Tanaka}} \bibnamefont{and}
  \bibinfo{author}{\bibfnamefont{H.}~\bibnamefont{Suzuki}},
  \bibinfo{journal}{Carbon} \textbf{\bibinfo{volume}{10}}, \bibinfo{pages}{253}
  (\bibinfo{year}{1972}), ISSN \bibinfo{issn}{0008-6223},
  \urlprefix\url{http://www.sciencedirect.com/science/article/pii/0008622372903235}.

\bibitem[{\citenamefont{Issi et~al.}(1983)\citenamefont{Issi, Heremans, and
  Dresselhaus}}]{issi_electronic_1983}
\bibinfo{author}{\bibfnamefont{J.~P.} \bibnamefont{Issi}},
  \bibinfo{author}{\bibfnamefont{J.}~\bibnamefont{Heremans}}, \bibnamefont{and}
  \bibinfo{author}{\bibfnamefont{M.~S.} \bibnamefont{Dresselhaus}},
  \bibinfo{journal}{Physical Review B} \textbf{\bibinfo{volume}{27}},
  \bibinfo{pages}{1333} (\bibinfo{year}{1983}),
  \urlprefix\url{http://link.aps.org/doi/10.1103/PhysRevB.27.1333}.

\bibitem[{\citenamefont{Heremans et~al.}(1981)\citenamefont{Heremans, Issi,
  Zabala-Martinez, Shayegan, and Dresselhaus}}]{heremans_thermoelectric_1981}
\bibinfo{author}{\bibfnamefont{J.}~\bibnamefont{Heremans}},
  \bibinfo{author}{\bibfnamefont{J.~P.} \bibnamefont{Issi}},
  \bibinfo{author}{\bibfnamefont{I.}~\bibnamefont{Zabala-Martinez}},
  \bibinfo{author}{\bibfnamefont{M.}~\bibnamefont{Shayegan}}, \bibnamefont{and}
  \bibinfo{author}{\bibfnamefont{M.~S.} \bibnamefont{Dresselhaus}},
  \bibinfo{journal}{Physics Letters A} \textbf{\bibinfo{volume}{84}},
  \bibinfo{pages}{387} (\bibinfo{year}{1981}), ISSN \bibinfo{issn}{0375-9601},
  \urlprefix\url{http://www.sciencedirect.com/science/article/pii/0375960181902176}.

\bibitem[{\citenamefont{Boxus et~al.}(1981)\citenamefont{Boxus, Poulaert, Issi,
  Mazurek, and Dresselhaus}}]{boxus_low_1981}
\bibinfo{author}{\bibfnamefont{J.}~\bibnamefont{Boxus}},
  \bibinfo{author}{\bibfnamefont{B.}~\bibnamefont{Poulaert}},
  \bibinfo{author}{\bibfnamefont{J.-P.} \bibnamefont{Issi}},
  \bibinfo{author}{\bibfnamefont{H.}~\bibnamefont{Mazurek}}, \bibnamefont{and}
  \bibinfo{author}{\bibfnamefont{M.~S.} \bibnamefont{Dresselhaus}},
  \bibinfo{journal}{Solid State Communications} \textbf{\bibinfo{volume}{38}},
  \bibinfo{pages}{1117} (\bibinfo{year}{1981}), ISSN \bibinfo{issn}{0038-1098},
  \urlprefix\url{http://www.sciencedirect.com/science/article/pii/0038109881909698}.

\bibitem[{\citenamefont{Dresselhaus and
  Dresselhaus}(1981)}]{dresselhaus_intercalation_1981}
\bibinfo{author}{\bibfnamefont{M.}~\bibnamefont{Dresselhaus}} \bibnamefont{and}
  \bibinfo{author}{\bibfnamefont{G.}~\bibnamefont{Dresselhaus}},
  \bibinfo{journal}{Advances in Physics} \textbf{\bibinfo{volume}{30}},
  \bibinfo{pages}{139} (\bibinfo{year}{1981}), ISSN \bibinfo{issn}{0001-8732},
  \urlprefix\url{http://www.tandfonline.com/doi/abs/10.1080/00018738100101367}.

\bibitem[{\citenamefont{Chiritescu et~al.}(2008)\citenamefont{Chiritescu,
  Cahill, Heideman, Lin, Mortensen, Nguyen, Johnson, Rostek, and
  B{\"o}ttner}}]{chiritescu_low_2008}
\bibinfo{author}{\bibfnamefont{C.}~\bibnamefont{Chiritescu}},
  \bibinfo{author}{\bibfnamefont{D.~G.} \bibnamefont{Cahill}},
  \bibinfo{author}{\bibfnamefont{C.}~\bibnamefont{Heideman}},
  \bibinfo{author}{\bibfnamefont{Q.}~\bibnamefont{Lin}},
  \bibinfo{author}{\bibfnamefont{C.}~\bibnamefont{Mortensen}},
  \bibinfo{author}{\bibfnamefont{N.~T.} \bibnamefont{Nguyen}},
  \bibinfo{author}{\bibfnamefont{D.}~\bibnamefont{Johnson}},
  \bibinfo{author}{\bibfnamefont{R.}~\bibnamefont{Rostek}}, \bibnamefont{and}
  \bibinfo{author}{\bibfnamefont{H.}~\bibnamefont{B{\"o}ttner}},
  \bibinfo{journal}{Journal of Applied Physics} \textbf{\bibinfo{volume}{104}},
  \bibinfo{pages}{033533} (\bibinfo{year}{2008}), ISSN
  \bibinfo{issn}{0021-8979, 1089-7550},
  \urlprefix\url{http://scitation.aip.org/content/aip/journal/jap/104/3/10.1063/1.2967722}.

\bibitem[{\citenamefont{Hsieh et~al.}(2009)\citenamefont{Hsieh, Chen, Li,
  Keblinski, and Cahill}}]{hsieh_pressure_2009}
\bibinfo{author}{\bibfnamefont{W.-P.} \bibnamefont{Hsieh}},
  \bibinfo{author}{\bibfnamefont{B.}~\bibnamefont{Chen}},
  \bibinfo{author}{\bibfnamefont{J.}~\bibnamefont{Li}},
  \bibinfo{author}{\bibfnamefont{P.}~\bibnamefont{Keblinski}},
  \bibnamefont{and} \bibinfo{author}{\bibfnamefont{D.~G.}
  \bibnamefont{Cahill}}, \bibinfo{journal}{Physical Review B}
  \textbf{\bibinfo{volume}{80}}, \bibinfo{pages}{180302}
  (\bibinfo{year}{2009}),
  \urlprefix\url{http://link.aps.org/doi/10.1103/PhysRevB.80.180302}.

\bibitem[{\citenamefont{Klemens and Pedraza}(1994)}]{klemens_thermal_1994}
\bibinfo{author}{\bibfnamefont{P.~G.} \bibnamefont{Klemens}} \bibnamefont{and}
  \bibinfo{author}{\bibfnamefont{D.~F.} \bibnamefont{Pedraza}},
  \bibinfo{journal}{Carbon} \textbf{\bibinfo{volume}{32}}, \bibinfo{pages}{735}
  (\bibinfo{year}{1994}), ISSN \bibinfo{issn}{0008-6223},
  \urlprefix\url{http://www.sciencedirect.com/science/article/pii/0008622394900965}.

\bibitem[{\citenamefont{Chen et~al.}(2013)\citenamefont{Chen, Wei, Chen, and
  Dames}}]{chen_anisotropic_2013}
\bibinfo{author}{\bibfnamefont{Z.}~\bibnamefont{Chen}},
  \bibinfo{author}{\bibfnamefont{Z.}~\bibnamefont{Wei}},
  \bibinfo{author}{\bibfnamefont{Y.}~\bibnamefont{Chen}}, \bibnamefont{and}
  \bibinfo{author}{\bibfnamefont{C.}~\bibnamefont{Dames}},
  \bibinfo{journal}{Physical Review B} \textbf{\bibinfo{volume}{87}},
  \bibinfo{pages}{125426} (\bibinfo{year}{2013}),
  \urlprefix\url{http://link.aps.org/doi/10.1103/PhysRevB.87.125426}.

\bibitem[{\citenamefont{Duda et~al.}(2009)\citenamefont{Duda, Smoyer, Norris,
  and Hopkins}}]{duda_extension_2009}
\bibinfo{author}{\bibfnamefont{J.~C.} \bibnamefont{Duda}},
  \bibinfo{author}{\bibfnamefont{J.~L.} \bibnamefont{Smoyer}},
  \bibinfo{author}{\bibfnamefont{P.~M.} \bibnamefont{Norris}},
  \bibnamefont{and} \bibinfo{author}{\bibfnamefont{P.~E.}
  \bibnamefont{Hopkins}}, \bibinfo{journal}{Applied Physics Letters}
  \textbf{\bibinfo{volume}{95}}, \bibinfo{pages}{031912}
  (\bibinfo{year}{2009}), ISSN \bibinfo{issn}{0003-6951, 1077-3118},
  \urlprefix\url{http://scitation.aip.org/content/aip/journal/apl/95/3/10.1063/1.3189087}.

\bibitem[{\citenamefont{Prasher}(2008)}]{prasher_thermal_2008}
\bibinfo{author}{\bibfnamefont{R.}~\bibnamefont{Prasher}},
  \bibinfo{journal}{Physical Review B} \textbf{\bibinfo{volume}{77}},
  \bibinfo{pages}{075424} (\bibinfo{year}{2008}),
  \urlprefix\url{http://link.aps.org/doi/10.1103/PhysRevB.77.075424}.

\bibitem[{\citenamefont{Wei et~al.}(2013)\citenamefont{Wei, Chen, and
  Dames}}]{wei_negative_2013}
\bibinfo{author}{\bibfnamefont{Z.}~\bibnamefont{Wei}},
  \bibinfo{author}{\bibfnamefont{Y.}~\bibnamefont{Chen}}, \bibnamefont{and}
  \bibinfo{author}{\bibfnamefont{C.}~\bibnamefont{Dames}},
  \bibinfo{journal}{Applied Physics Letters} \textbf{\bibinfo{volume}{102}},
  \bibinfo{pages}{011901} (\bibinfo{year}{2013}), ISSN
  \bibinfo{issn}{00036951},
  \urlprefix\url{http://apl.aip.org.clsproxy.library.caltech.edu/resource/1/applab/v102/i1/p011901_s1}.

\bibitem[{\citenamefont{Wei et~al.}(2014)\citenamefont{Wei, Yang, Chen, Bi, Li,
  and Chen}}]{wei_phonon_2014}
\bibinfo{author}{\bibfnamefont{Z.}~\bibnamefont{Wei}},
  \bibinfo{author}{\bibfnamefont{J.}~\bibnamefont{Yang}},
  \bibinfo{author}{\bibfnamefont{W.}~\bibnamefont{Chen}},
  \bibinfo{author}{\bibfnamefont{K.}~\bibnamefont{Bi}},
  \bibinfo{author}{\bibfnamefont{D.}~\bibnamefont{Li}}, \bibnamefont{and}
  \bibinfo{author}{\bibfnamefont{Y.}~\bibnamefont{Chen}},
  \bibinfo{journal}{Applied Physics Letters} \textbf{\bibinfo{volume}{104}},
  \bibinfo{pages}{081903} (\bibinfo{year}{2014}), ISSN
  \bibinfo{issn}{0003-6951, 1077-3118},
  \urlprefix\url{http://scitation.aip.org/content/aip/journal/apl/104/8/10.1063/1.4866416}.

\bibitem[{\citenamefont{Yang et~al.}(2014)\citenamefont{Yang, Shen, Yang,
  Evans, Wei, Chen, Zinn, Chen, Prasher, Xu et~al.}}]{yang_phonon_2014}
\bibinfo{author}{\bibfnamefont{J.}~\bibnamefont{Yang}},
  \bibinfo{author}{\bibfnamefont{M.}~\bibnamefont{Shen}},
  \bibinfo{author}{\bibfnamefont{Y.}~\bibnamefont{Yang}},
  \bibinfo{author}{\bibfnamefont{W.~J.} \bibnamefont{Evans}},
  \bibinfo{author}{\bibfnamefont{Z.}~\bibnamefont{Wei}},
  \bibinfo{author}{\bibfnamefont{W.}~\bibnamefont{Chen}},
  \bibinfo{author}{\bibfnamefont{A.~A.} \bibnamefont{Zinn}},
  \bibinfo{author}{\bibfnamefont{Y.}~\bibnamefont{Chen}},
  \bibinfo{author}{\bibfnamefont{R.}~\bibnamefont{Prasher}},
  \bibinfo{author}{\bibfnamefont{T.~T.} \bibnamefont{Xu}},
  \bibnamefont{et~al.}, \bibinfo{journal}{Physical Review Letters}
  \textbf{\bibinfo{volume}{112}}, \bibinfo{pages}{205901}
  (\bibinfo{year}{2014}),
  \urlprefix\url{http://link.aps.org/doi/10.1103/PhysRevLett.112.205901}.

\bibitem[{\citenamefont{Harb et~al.}(2012)\citenamefont{Harb, von
  Korff~Schmising, Enquist, Jurgilaitis, Maximov, Shvets, Obraztsov, Khakhulin,
  Wulff, and Larsson}}]{harb_c-axis_2012}
\bibinfo{author}{\bibfnamefont{M.}~\bibnamefont{Harb}},
  \bibinfo{author}{\bibfnamefont{C.}~\bibnamefont{von Korff~Schmising}},
  \bibinfo{author}{\bibfnamefont{H.}~\bibnamefont{Enquist}},
  \bibinfo{author}{\bibfnamefont{A.}~\bibnamefont{Jurgilaitis}},
  \bibinfo{author}{\bibfnamefont{I.}~\bibnamefont{Maximov}},
  \bibinfo{author}{\bibfnamefont{P.~V.} \bibnamefont{Shvets}},
  \bibinfo{author}{\bibfnamefont{A.~N.} \bibnamefont{Obraztsov}},
  \bibinfo{author}{\bibfnamefont{D.}~\bibnamefont{Khakhulin}},
  \bibinfo{author}{\bibfnamefont{M.}~\bibnamefont{Wulff}}, \bibnamefont{and}
  \bibinfo{author}{\bibfnamefont{J.}~\bibnamefont{Larsson}},
  \bibinfo{journal}{Applied Physics Letters} \textbf{\bibinfo{volume}{101}},
  \bibinfo{pages}{233108} (\bibinfo{year}{2012}), ISSN
  \bibinfo{issn}{00036951},
  \urlprefix\url{http://apl.aip.org.clsproxy.library.caltech.edu/resource/1/applab/v101/i23/p233108_s1}.

\bibitem[{\citenamefont{Chen}(2005)}]{GangBook}
\bibinfo{author}{\bibfnamefont{G.}~\bibnamefont{Chen}},
  \emph{\bibinfo{title}{Nanoscale Energy Transport and Conversion}}
  (\bibinfo{publisher}{Oxford University Press}, \bibinfo{address}{New York},
  \bibinfo{year}{2005}).

\bibitem[{\citenamefont{Hua and Minnich}(2014)}]{Hua:2014}
\bibinfo{author}{\bibfnamefont{C.}~\bibnamefont{Hua}} \bibnamefont{and}
  \bibinfo{author}{\bibfnamefont{A.~J.} \bibnamefont{Minnich}},
  \bibinfo{journal}{Physical Review B} \textbf{\bibinfo{volume}{89}},
  \bibinfo{pages}{094302} (\bibinfo{year}{2014}),
  \urlprefix\url{http://link.aps.org/doi/10.1103/PhysRevB.89.094302}.

\bibitem[{\citenamefont{Vermeersch et~al.}(2014)\citenamefont{Vermeersch,
  Carrete, Mingo, and Shakouri}}]{vermeersch_superdiffusive_2014}
\bibinfo{author}{\bibfnamefont{B.}~\bibnamefont{Vermeersch}},
  \bibinfo{author}{\bibfnamefont{J.}~\bibnamefont{Carrete}},
  \bibinfo{author}{\bibfnamefont{N.}~\bibnamefont{Mingo}}, \bibnamefont{and}
  \bibinfo{author}{\bibfnamefont{A.}~\bibnamefont{Shakouri}},
  \bibinfo{journal}{{arXiv}:1406.7341 [cond-mat]}  (\bibinfo{year}{2014}),
  \bibinfo{note}{{arXiv}: 1406.7341},
  \urlprefix\url{http://arxiv.org/abs/1406.7341}.

\bibitem[{\citenamefont{Johnson et~al.}(2013)\citenamefont{Johnson, Maznev,
  Cuffe, Eliason, Minnich, Kehoe, Torres, Chen, and
  Nelson}}]{johnson_direct_2013}
\bibinfo{author}{\bibfnamefont{J.~A.} \bibnamefont{Johnson}},
  \bibinfo{author}{\bibfnamefont{A.~A.} \bibnamefont{Maznev}},
  \bibinfo{author}{\bibfnamefont{J.}~\bibnamefont{Cuffe}},
  \bibinfo{author}{\bibfnamefont{J.~K.} \bibnamefont{Eliason}},
  \bibinfo{author}{\bibfnamefont{A.~J.} \bibnamefont{Minnich}},
  \bibinfo{author}{\bibfnamefont{T.}~\bibnamefont{Kehoe}},
  \bibinfo{author}{\bibfnamefont{C.~M.~S.} \bibnamefont{Torres}},
  \bibinfo{author}{\bibfnamefont{G.}~\bibnamefont{Chen}}, \bibnamefont{and}
  \bibinfo{author}{\bibfnamefont{K.~A.} \bibnamefont{Nelson}},
  \bibinfo{journal}{Physical Review Letters} \textbf{\bibinfo{volume}{110}},
  \bibinfo{pages}{025901} (\bibinfo{year}{2013}),
  \urlprefix\url{http://link.aps.org/doi/10.1103/PhysRevLett.110.025901}.

\bibitem[{\citenamefont{Minnich et~al.}(2009)\citenamefont{Minnich,
  Dresselhaus, Ren, and Chen}}]{AustinEES}
\bibinfo{author}{\bibfnamefont{A.~J.} \bibnamefont{Minnich}},
  \bibinfo{author}{\bibfnamefont{M.~S.} \bibnamefont{Dresselhaus}},
  \bibinfo{author}{\bibfnamefont{Z.~F.} \bibnamefont{Ren}}, \bibnamefont{and}
  \bibinfo{author}{\bibfnamefont{G.}~\bibnamefont{Chen}},
  \bibinfo{journal}{Energy \& Environmental Science}
  \textbf{\bibinfo{volume}{2}}, \bibinfo{pages}{466 } (\bibinfo{year}{2009}),
  \bibinfo{note}{(Invited, peer-reviewed. Among the top 10 downloaded articles
  in EES for July, August, December 2009, January-May, July-December 2010,
  January 2011).}

\bibitem[{\citenamefont{Gradshteyn and Ryzhik}(1980)}]{Rzhik}
\bibinfo{author}{\bibfnamefont{I.~S.} \bibnamefont{Gradshteyn}}
  \bibnamefont{and} \bibinfo{author}{\bibfnamefont{I.~M.}
  \bibnamefont{Ryzhik}}, \emph{\bibinfo{title}{Table of integrals, series and
  products}} (\bibinfo{publisher}{New York: Academic Press},
  \bibinfo{year}{1980}).

\bibitem[{\citenamefont{Minnich}(2012)}]{minnich_determining_2012}
\bibinfo{author}{\bibfnamefont{A.~J.} \bibnamefont{Minnich}},
  \bibinfo{journal}{Physical Review Letters} \textbf{\bibinfo{volume}{109}},
  \bibinfo{pages}{205901} (\bibinfo{year}{2012}),
  \urlprefix\url{http://link.aps.org/doi/10.1103/PhysRevLett.109.205901}.

\bibitem[{\citenamefont{Grant and Boyd}(2008)}]{gb08}
\bibinfo{author}{\bibfnamefont{M.}~\bibnamefont{Grant}} \bibnamefont{and}
  \bibinfo{author}{\bibfnamefont{S.}~\bibnamefont{Boyd}}, in
  \emph{\bibinfo{booktitle}{Recent Advances in Learning and Control}}, edited
  by \bibinfo{editor}{\bibfnamefont{V.}~\bibnamefont{Blondel}},
  \bibinfo{editor}{\bibfnamefont{S.}~\bibnamefont{Boyd}}, \bibnamefont{and}
  \bibinfo{editor}{\bibfnamefont{H.}~\bibnamefont{Kimura}}
  (\bibinfo{publisher}{Springer-Verlag Limited}, \bibinfo{year}{2008}), Lecture
  Notes in Control and Information Sciences, pp. \bibinfo{pages}{95--110},
  \bibinfo{note}{\url{http://stanford.edu/~boyd/graph_dcp.html}}.

\bibitem[{\citenamefont{Grant and Boyd}(2011)}]{cvx}
\bibinfo{author}{\bibfnamefont{M.}~\bibnamefont{Grant}} \bibnamefont{and}
  \bibinfo{author}{\bibfnamefont{S.}~\bibnamefont{Boyd}},
  \emph{\bibinfo{title}{{CVX}: Matlab software for disciplined convex
  programming, version 1.21}}, \bibinfo{howpublished}{\url{../../cvx}}
  (\bibinfo{year}{2011}).

\end{thebibliography}

\end{document}